\documentclass[aps,english,superscriptaddress,preprintnumbers,reprint,footinbib,amsmath,amssymb,prb]{revtex4-2}
\usepackage[version=3]{mhchem}

\usepackage{graphicx}
\usepackage{dcolumn}
\usepackage{bm}
\usepackage{hyperref}
\hypersetup{
    colorlinks=true,
    linkcolor=blue,
    filecolor=blue,
    urlcolor=blue,
   citecolor=blue,
}
\usepackage{enumitem}
\usepackage{siunitx}
\usepackage{times} 
\usepackage{newtxtext,newtxmath}

\begin{document}

\title{Accessing the degree of Majorana nonlocality in a quantum dot-optical microcavity system}

\author{L. S. Ricco}
\affiliation{Science Institute, University of Iceland, Dunhagi-3, IS-107, Reykjavik, Iceland\\ (corresponding author: lsricco@hi.is)}

\author{V. K. Kozin}
\affiliation{Science Institute, University of Iceland, Dunhagi-3, IS-107, Reykjavik, Iceland}
\affiliation{Department of Physics, ITMO University, St.~Petersburg 197101, Russia}

\author{A. C. Seridonio}
\affiliation{S\~ao Paulo State University (Unesp), School of Engineering, Department of Physics and Chemistry, 15385-000, Ilha Solteira-SP, Brazil}
\affiliation{S\~ao Paulo State University (Unesp), IGCE, Department of Physics, 13506-970, Rio Claro-SP, Brazil}

\author{I. A. Shelykh}
\affiliation{Science Institute, University of Iceland, Dunhagi-3, IS-107, Reykjavik, Iceland}
\affiliation{Department of Physics, ITMO University, St.~Petersburg 197101, Russia}

\date{\today}

\begin{abstract}
We explore the tunneling transport properties of a quantum dot embedded in an optical microcavity and coupled to a semiconductor-superconductor one-dimensional nanowire (Majorana nanowire) hosting Majorana zero modes (MZMs) at their edges. Conductance profiles reveal that strong light-matter coupling can be employed to distinguish between the cases of highly nonlocal MZMs, overlapped MZMs and MZMs with less degree of nonlocal feature. Moreover, we show that it is possible to access the degree of Majorana nonlocality (topological quality factor) by changing the dot spectrum through photon-induced transitions tuned by an external pump applied to the microcavity.
\end{abstract}

\maketitle
\textbf{Introduction}

Over the past decade, a huge effort in both theoretical and experimental fields has been performing in the quest for an unquestionable signature of exotic `half-fermionic' states, the so-called Majorana zero-modes (MZMs) in quasi-one dimensional hybrid semiconductor-superconductor nanowires with strong spin-orbit coupling subject to external magnetic field, termed as Majorana nanowires~\cite{RevMajoranaAlicea,RevMajoranaFranz,RevMajoranaAguado,LutchynReviewMat2018,JelenaReviewJAP2021}. In such devices, each isolated `half-fermionic' MZM appears at one of the opposite nanowire ends when the bulk of the system undergoes a topological phase transition~\cite{LutchynPRL2010,OregPRL2010,Zhang2019}.

In tunneling spectroscopy measurements performed through a Majorana nanowire~\cite{MourikScience2012,KrogstrupNatMater2015,AlbrechtNature2009,DengScience2016,DengPRB2018,ZhangNatNanotech2018,LutchynReviewMat2018,Zhang2019,zhang2021large}, the emergence of a quantized zero-bias peak (ZBP) robust to changing of relevant system parameters such as magnetic field and gate voltages is considered as a strong evidence supporting the emergence of the isolated MZMs. However, it has been extensively demonstrated that other physical mechanisms such as the formation of trivial zero-energy Andreev bound state (ABS) at wire ends due to inhomogeneous smooth confining potentials~\cite{LiuPRB2017,LiuPRB2018,HellPRB2018,LaiPRB2019,MoorePRB2018,Vuik2019,Pan2020,prada2019andreev,PanGenericQuantized2020,pan2021quantized} and disorder-induced bound states~\cite{PikulinNJP2012,Pan2020,pan2020disorder,pan2021quantized} can produce a robust  quantized ZBP, leading to an ambiguity of the MZM signature. 

The emergence of a trivial zero-energy ABS can be mathematically described as resulting from two half-fermionic states with some spatial separation between them. In this scenario, a quantized ZBP may arise if one of these half-fermionic states couples to a tunneling spectroscopy probe stronger than the other~\cite{MoorePRB2018,Vuik2019,zhang2021large}. This kind of trivial ABS formed by two half-fermionic states with small spatial separation between them has been dubbed as partially separated ABS (ps-ABS)~\cite{MoorePRB2018} or quasi-Majoranas (quasi-MZM)~\cite{Vuik2019}. Despite some recent advances, distinguishing ZBPs resulting from genuine topological MZMs, trivial quasi-MZMs and disorder still remains a challenge~\cite{Pan2020,PanGenericQuantized2020,prada2019andreev,zhang2021large,pan2021quantized}.

\begin{figure}[t]
	\centerline{\includegraphics[width=3.5in,keepaspectratio]{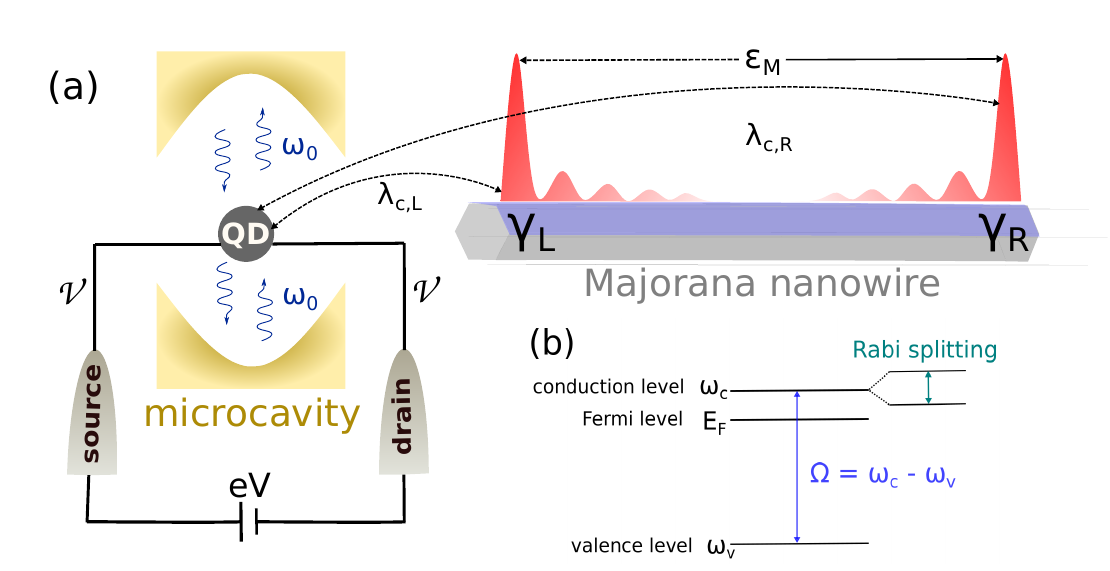}}
	\caption{\label{fig:System}(a) The sketch of the considered system: A  quantum dot (QD) embedded inside a single-mode optical cavity with frequency $\omega_{0}$. An electron in the conduction level of the dot couples with both the left ($\gamma_{R}$) and the right ($\gamma_{L}$) Majorana zero-modes (MZMs) located at the opposite
		ends of the Majorana nanowire, with strengths $\lambda_{c,L}$ and $\lambda_{c,R}$, respectively. The overlap between the wavefunctions of MZMs is given by $\varepsilon_M$. The tunneling conductance through the QD can by probed by means of source and 
		drain metallic leads symmetrically coupled ($\mathcal{V}$) to the dot. The energy of the cavity photons is brought in resonance with interband transition in the dot $\Omega$. (b) The scheme of a QD energy levels: the valence level $\omega_{v}$ is far below the Fermi level $E_{F}=0$, while the conduction level $\omega_{c}$ is above $E_{F}$. Strong coupling with cavity photons results in a Rabi splitting of the levels $\Omega_R$, which is determined by the oscillator strength of the optical transition and the geometry of the cavity. }
\end{figure}

From the perspective of Majorana-based qubits~\cite{AesenPhysRevX6031016(2016)}, the possibility of performing decoherence-free quantum computing operations lies on the ability of storing the information nonlocaly~\cite{Kitaev2001,RevNonabelian2008,Vuik2019,Penaranda2018}, once each MZM is far apart from each other for the ideal situation of longer and pristine nanowires. Hereupon, it is crucial to obtain the information about the spatial localization of the wave functions corresponding to these MZMs. This can be achieved by measuring the so-called degree of Majorana nonlocality~\cite{Prada2017,Penaranda2018}, which indirectly quantifies the information about the localization of the MZMs wavefunctions along the nanowire. This quantity can be accessed by means of a quantum dot (QD) working as a local tunneling spectroscopy probe at one of the ends of a Majorana nanowire~\cite{Baranger2011,VernekPRB2014,DengScience2016,DengPRB2018,SciRepIsoconductance2021}. The degree of Majorana nonlocality is defined as $\eta^{2} = (\lambda_{c,R}/\lambda_{c,L})$ for weakly overlapped MZMs ($\varepsilon_{M}\ll \lambda_{c,R}$),  where $\lambda_{c,L}$ and $\lambda_{c,R}$ are the coupling between the dot and the left and right MZMs, respectively [see Fig.~\ref{fig:System}(a)]. Equivalently, this quantity provides a topological quality factor $q = 1 - (\lambda_{c,R}/\lambda_{c,L})$ of the nanowire~\cite{DJClarcke2017,RiccoPRB2020}. Highly nonlocal MZMs are characterized by $\eta \rightarrow 0$ ($q \rightarrow 1$), while $\eta \rightarrow 1$ ($q \rightarrow 0$) indicates MZMs with corresponding wavefunctions not well-localized at the nanowire ends. Experimentally~\cite{Prada2017,DengPRB2018}, the degree of Majorana nonlocality (quality factor) can be accessed by measuring the corresponding energies of anticrossing patterns appearing in tunneling conductance profiles as functions of both applied bias-voltage through the QD-nanowire and QD gate-voltage.

In the present work, we show that light-matter coupling can be employed as an alternative to probe the degree of Majorana nonlocality  through tunneling conductance experiments. We consider the system schematically shown in Fig.~\ref{fig:System}(a): a QD placed inside a single-mode optical cavity with frequency $\omega_{0}$, which is tuned in resonance with valence-to-conduction band optical transition, i.e, $\Omega =\omega_{0}$, see Fig.~\ref{fig:System}(b).
The conductance through the dot can be accessed by metallic source-drain leads. Moreover, the QD couples with both left and right MZMs hosted at the edges of the Majorana nanowire. We demonstrate, that conductance profiles through the QD as functions of both bias voltage and mean photon occupation adjusted by external pump reveal distinct patterns for the cases of highly nonlocal MZMs, overlapped MZMs and not so well localized MZMs. We also show that it is possible to access the degree of Majorana nonlocality (topological quality factor) in conductance profiles by tuning the mean photon occupation instead of the QD gate-voltage as in the Prada´s original proposal~\cite{Prada2017,Penaranda2018,DJClarcke2017}.

To the best of our knowledge, experimental setups which combine optical and electrical methods to explore the underlying physics of MZMs have not been considered previously. However, QDs in microcavity quantum electrodynamics (QED) devices are routinely studied in optical experiments~\cite{qd_in_cav_ref1, qd_in_cav_ref2} and, in recent years, some theoretical works have employed cavity QED for studying MZMs~\cite{MirceaPRL2012,DartiailhPRL2017,PhysRevResearch.2.043264,Contamin2021condmat}. In such theoretical proposals, the whole Majorana nanowire is placed in a (microwave) cavity, whereas in our approach only the QD is sandwiched between the mirrors that form an optical microcavity. Moreover, in our theoretical approach we introduce a new (optical) degree of freedom in the system such that, on the one hand, it affects only a small part of the full system in contrary to the previous approaches, and on the other hand, allows one optical manipulation of the electrical conductance, thus providing an alternative to access  degree of Majorana nonlocality via optical means. It's worthy noticing that the bridging between cavity QED and topological superconductors supporting Majorana excitations not only opens new possibilities for MZMs detection schemes, but also can pave a new way for performing quantum computing operations with photons~\cite{Contamin2021condmat,Barz_2015,Science8770(2020)} in Majorana based qubit schemes.


\textbf{Results and Discussion}

In what follows, we analyze the conductance through the QD [Eq.~(\ref{eq:Conductance_T=0_only_Ec})] as a function of bias-voltage $eV$, in presence of a photonic field in the cavity [Fig.~\ref{fig:System}(a)]. We take the effective broadening $\Gamma=\SI{40}{\micro\electronvolt}$ as the energy unit~\cite{VernekPRB2014,baranski2020dynamical}. We also use typical parameters for QDs embedded in microcavities~\cite{Khitrova2006}, taking $\omega_v \ll E_F$, $\omega_c = 100\Gamma\approx \SI{968}{\giga\hertz}$, $\Omega_R = 2.5\Gamma=\SI{0.1}{\milli\electronvolt}$ ($\approx \SI{24}{\giga\hertz}$). The energy of a cavity mode is taken in resonance with the excitonic transition in the QD, $\omega_0 = \Omega=35.1\times10^{3}\Gamma\approx \SI{1.4}{\electronvolt}$ ($\SI{338}{\tera\hertz}$).
\begin{figure*}[t]
	\centerline{\includegraphics[width=6.6in,keepaspectratio]{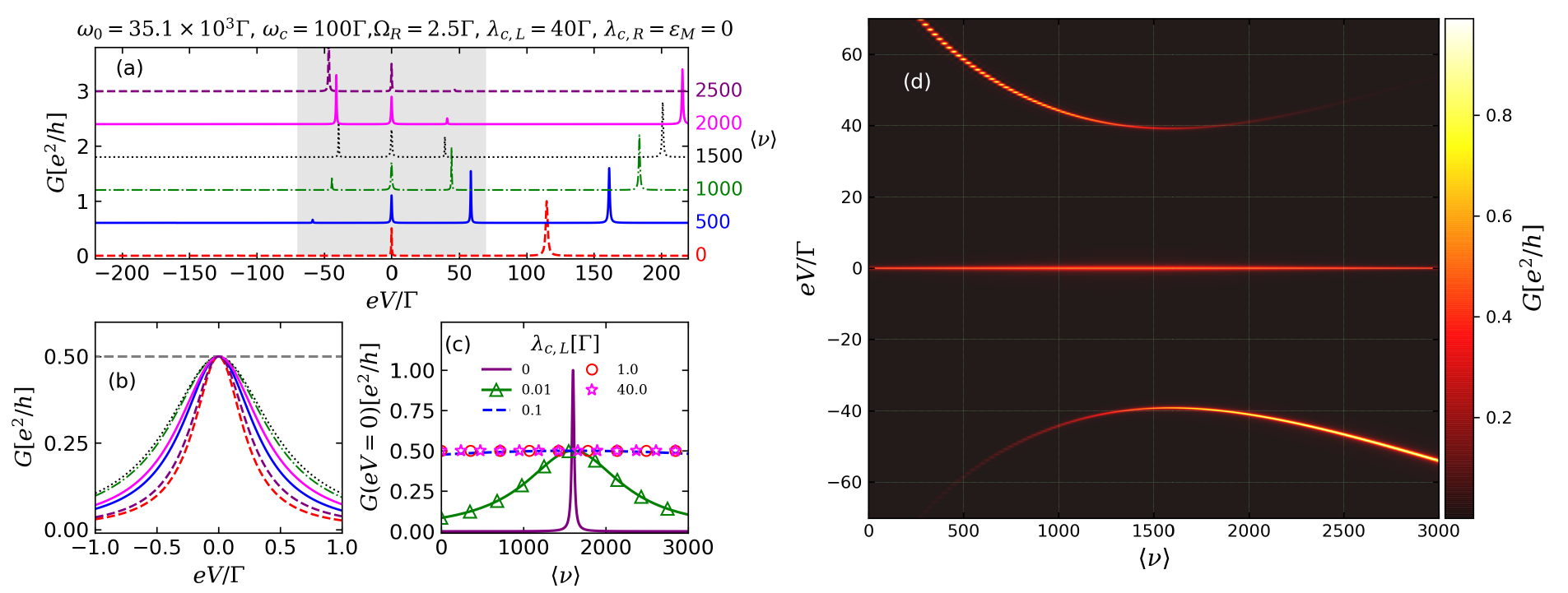}}
	\caption{ (a) Conductance through the QD [Eq.~(\ref{eq:Conductance_T=0_only_Ec})] as a function of bias-voltage $eV$ describing the case of highly nonlocal MZMs ($\varepsilon_{M}=\lambda_{c,R}=0$, $\eta = 0$) for increasing number of the excitations in the system $\langle\nu\rangle$, which can be tuned by external pump (the curves are offset along the $y$ axis for the better viewing). One clearly sees the appearance of additional peaks due to the Rabi splitting. (b) Zoom of the conductance around $eV=0$. (c) Conductance through the QD at zero-bias $eV = 0$ for increasing $\langle\nu\rangle$. (d) Colormap showing the conductance behavior as a function of both the $eV$ and $\langle \nu \rangle$, for a range of $eV$ corresponding to gray region of panel (a).    
	\label{fig:Result1}}
\end{figure*}

Fig.~\ref{fig:Result1} describes the conductance as a function of $eV$ when the QD is coupled only with the left MZM ($\lambda_{c,L}=40\Gamma$), corresponding to the situation of highly nonlocal MZMs ($\lambda_{c,R}=\varepsilon_{M}=0$, $\eta = 0$), for several values of $\langle \nu \rangle$. 
This ideal condition of true topological MZMs is expected for clean disorder-free nanowires~\cite{pan2020disorder,pan2021quantized}, long enough to avoid the overlap between the Majorana wavefunctions located at the opposite ends~\cite{prada2019andreev}. In absence of the cavity photons ($\langle \nu \rangle = 0$), a robust ZBP with $G(0)=0.5e^{2}/h$ appears, as it is shown by the red dashed lines in Figs.~\ref{fig:Result1}(a) and (b), characterizing the `half-fermionic' nature of an isolated MZM leaking into the QD~\cite{VernekPRB2014,TijerinaPRB2015,DengScience2016,baranski2020dynamical}.The conductance peak localized at $eV\approx 100\Gamma$ corresponds to the conduction level of the dot renormalized by the finite QD-left MZM coupling.

For a finite photon occupation $\langle\nu\rangle = 500$ [solid blue line, Fig~\ref{fig:Result1}(a)], optically-induced transitions between valence and conduction levels of the QD come into play, splitting the single peak associated with the QD conduction level into two polariton peaks located at $eV\approx 50\Gamma$ and $eV\approx 150\Gamma$. The higher is the value of $\langle\nu\rangle$, the bigger is the distance between the polariton peaks [Fig.~\ref{fig:Result1}(a), dash-dotted green and dotted black lines]. This behavior resembles the Rabi splitting for an individual dot inside a cavity, for which  $2\Omega_{R}\sqrt{\langle\nu\rangle}$~\cite{Cummings1965} [Fig.~\ref{fig:System}(b)]. Note, however, that for sufficiently high values of $\langle\nu\rangle$ additional polariton peaks stemming from indirect MZM to photon coupling of lower amplitude appear (see black and magenta curves). 

The amplitude of the peak at zero-bias remains unchanged  with increase of the number of the photons [Figs~\ref{fig:Result1}(b)]. This robustness is characteristic for the situation of highly-nonlocal MZMs~\cite{Prada2017,pan2021quantized,Pan2020,prada2019andreev}. In the same time, the width of the ZBP is monotonously increasing with increase of the number of the cavity photons, which means that the latter affect the effective lifetime of the electronic states of the dot and MZMs.  

In Fig.~\ref{fig:Result1}(c), the conductance behavior at zero-bias as a function of $\langle \nu \rangle$ is shown for several values of $\lambda_{c,L}$. A well-defined plateau of $0.5e^{2}/h$, independent on the mean photon occupation, characterizes the robustness of the ZBP for the cases of strong QD-nanowire coupling $\lambda_{c,L}\gtrsim \Gamma$ [magenta stars and red circles]. For smaller values of $\lambda_{c,L}$, the plateau is destroyed, as can be seen in the green line with triangles of Fig.~\ref{fig:Result1}(c), corresponding to $\lambda_{c,L}=10^{-2}\Gamma$. When the QD is totally decoupled from the Majorana nanowire [solid purple line, $\lambda_{c,L}=0$], the conductance at $eV=0$ exhibits a single resonance at $\langle \nu \rangle=1600$, which corresponds to the crossing between the photon-induced lower polariton peak and Fermi energy.           

The colormap of Fig.~\ref{fig:Result1}(d) summarizes the conductance behavior as a function of both $eV$ and $\langle \nu \rangle$, corresponding to a range of bias-voltage within the gray region of Fig.~\ref{fig:Result1}(a). One can notice that the photon-induced peak above $eV=0$ have its amplitude reduced with the co-emergence of a conductance peak symmetrically localized below $eV=0$ and a ZBP with unchanged $0.5e^2/h$ height pinned at zero-bias. Direct comparison between profile of Fig.~\ref{fig:Result1}(d) and the corresponding results for highly nonlocal MZMs found by Prada et al.~\cite{Prada2017} and others~\cite{DengScience2016,DJClarcke2017,DengPRB2018,RiccoPRB2019} shows that the increasing of the mean photon occupation (i.e. increasing the pump intensity) is a fast indirect way of tuning the QD energy without applying any gate-voltage due to renormalization of the QD spectrum in the strong light-matter coupling regime, which bridges Quantum Optics and the physics of MZMs.
\begin{figure*}[t]
	\centerline{\includegraphics[width=6.6in,keepaspectratio]{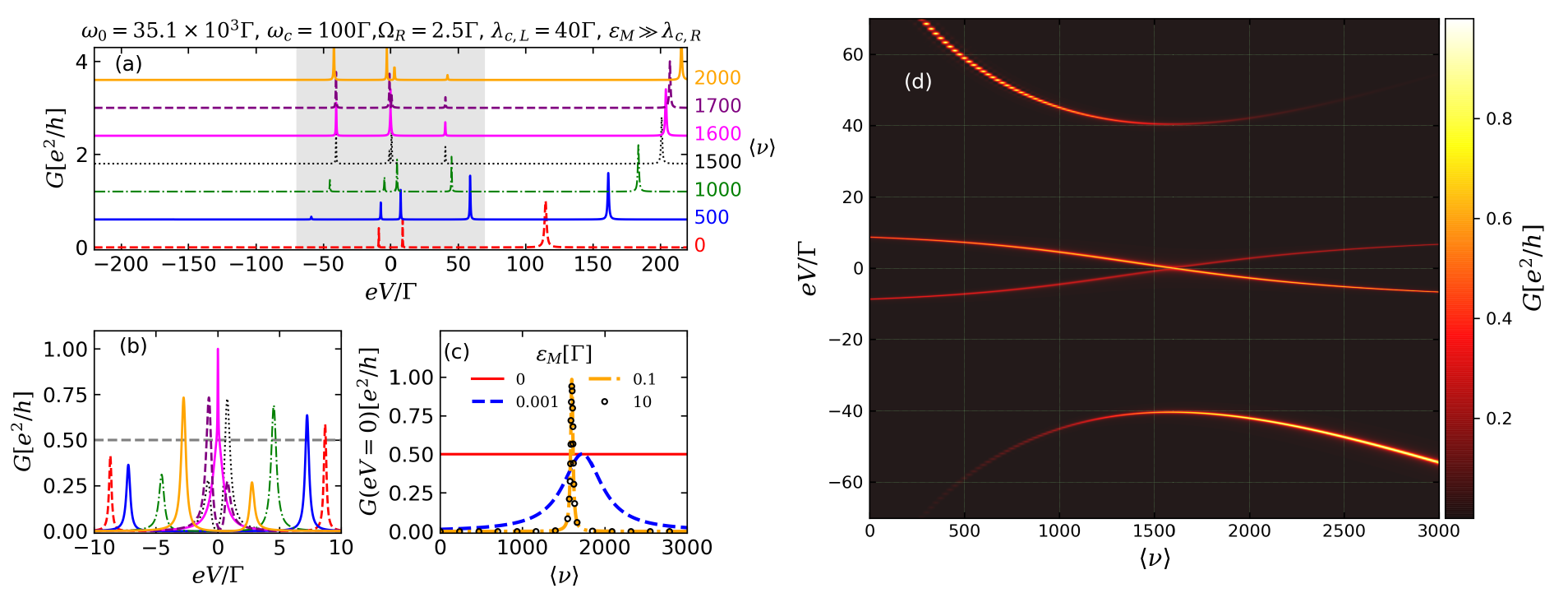}}
	\caption{ (a) Conductance through the QD [Eq.~(\ref{eq:Conductance_T=0_only_Ec})] as a function of bias-voltage $eV$ describing the case of overlapped MZMs localized at the nanowire ends ($\varepsilon_{M}=10\Gamma$, $\lambda_{c,R} = 0$) for increasing numbers of photon occupation $\langle\nu\rangle$. For a better viewing, the curves are offset along the $y$ axis. (b) Same curves depicted in panel (a), but considering $eV$ only near zero-bias without offset along $y$ axis. (c) Conductance through the QD at zero-bias $eV = 0$ for increasing number of $\langle \nu \rangle$. (d) Colormap showing the conductance behavior as a function of both the $eV$ and $\langle \nu \rangle$, for a range of $eV$ corresponding to gray region of panel (a).     
	\label{fig:Result2}}
\end{figure*}

\begin{figure*}[t]
	\centerline{\includegraphics[width=6.6in,keepaspectratio]{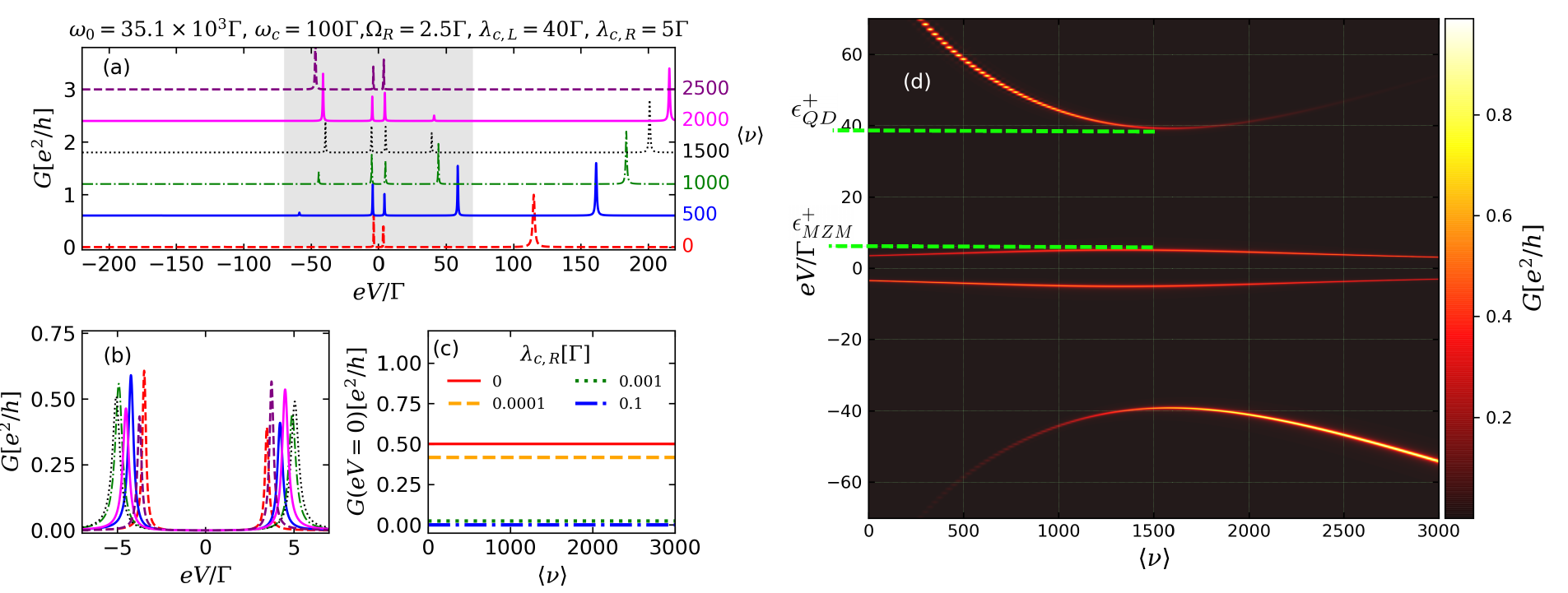}}
	\caption{(a) Conductance through the QD [Eq.~(\ref{eq:Conductance_T=0_only_Ec})] as a function of bias-voltage $eV$ describing the case of MZMs with less degree of nonlocality ($\lambda_{c,R} \gg \varepsilon_M$) for increasing number of photon occupation $\langle\nu\rangle$. For a better viewing, the curves are offset along the $y$ axis. (b) Same curves depicted in panel (a), but considering $eV$ only near zero-bias without offset along $y$ axis. (c) Conductance through the QD at zero-bias $eV = 0$ for increasing number of $\langle \nu \rangle$ considering distinct values of $\lambda_{c,R}$ . (d) Colormap showing the conductance behavior as a function of both the $eV$ and $\langle \nu \rangle$, for a range of $eV$ corresponding to gray region of panel (a). Dashed light-green lines indicate the energies $\epsilon_{QD}^{+}$ and $\epsilon_{MZMs}^{+}$ of anticrossing points employed to estimate the degree of Majorana nonlocality $\Omega_{M}^{2}=\epsilon_{MZMs}^{+}/\epsilon_{QD}^{+}$, as proposed by Prada et al.~\cite{Prada2017}. For the data shown in (d), $\Omega_{M}\approx 0.35$.
	\label{fig:Result3}}
\end{figure*}

It is worth noticing that the quantized ZBP Majorana amplitude of $0.5e^{2}/h$~\cite{Baranger2011,VernekPRB2014,TijerinaPRB2015} shown in the conductance profiles of Fig.~\ref{fig:Result1} is distinct from the $2e^{2}/h$ quantized Majorana conductance, expected for typical tunneling experiments with Majorana nanowires~\cite{Pan2020,pan2021quantized}. This difference comes from the underlying mechanism of electronic transport for each case. For characteristic setups of tunneling spectroscopy experiments~\cite{LutchynReviewMat2018,Zhang2019,zhang2021large}, the $2e^{2}/h$ quantized conductance is a result of a perfect resonant zero-energy Andreev reflection in the normal lead-Majorana nanowire interface~\cite{RevMajoranaAguado}. In our proposal otherwise [Fig.~\ref{fig:System}(a)], the transport occurs by normal electron tunneling through the QD due to an applied voltage $eV$ between the source-drain metallic leads, where the quantum of conductance is $e^{2}/h$ [Eq.~(\ref{eq:Conductance_T=0_only_Ec})] per spin channel. Hence, for an isolated MZM which has leaked into the QD level~\cite{VernekPRB2014,TijerinaPRB2015}, the amplitude of the ZBP is half of the quantum of conductance for an ordinary spinless electron, indicating the half-fermionic nature of a Majorana mode at zero-energy. 

Fig.~\ref{fig:Result2} shows the profiles of conductance through QD coupled to the Majorana nanowire for the case of MZMs still localized at opposite wire ends, but with  finite overlap between them ($\varepsilon_M = 10\Gamma\gg \lambda_{c,R}$), which may describe the situation of shorter nanowires~\cite{prada2019andreev}. For a null photon occupation, $G(eV)$ [Eq.~(\ref{eq:Conductance_T=0_only_Ec})] is characterized by two near zero-energy peaks at $eV\approx\pm\varepsilon_{M}$ [Figs.~\ref{fig:Result2}(a) and (b), dashed red lines] coming from the ZBP splitting due to the finite overlap~\cite{Baranger2011} and a third peak at $eV\approx 100\Gamma$, corresponding to the QD conduction level renormalized by the QD-nanowire coupling [Figs.~\ref{fig:Result2}(a), dashed red line]. As the mean photon occupation is increased, in analogy to what happened for nonoverlapped MZMs, the peak corresponding to the QD conduction level splits into two polariton peaks [Fig.~\ref{fig:Result2}(a), solid blue line], with the value of this splitting depending on $\langle \nu \rangle$. It can be seen that the increase of the mean photon occupation strongly affects the pattern of near zero-energy peaks associated with overlapped MZMs [see Fig.~\ref{fig:Result2}(b)]. First, they become closer to each other, and coalesce into a single peak at zero-bias with height of $e^{2}/h$ for $\langle \nu \rangle=1600$. Further increase of the number of cavity photons leads to the reappearance of the splitting. 

The conductance profile at $eV=0$ as a function of the mean photon occupation is shown in Fig.~\ref{fig:Result2}(c). One can notice that the zero-bias conductance plateau corresponding to nonverlapped and highly nonlocal MZMs [solid red line] is destroyed even by a small overlap [dashed blue line]. For bigger values of the overlap strength $\varepsilon_M$, $G(eV=0)$ as a function of $\langle \nu \rangle$, exhibits a single peak at $\langle \nu \rangle=1600$, showing the behavior similar to those corresponding to a QD embedded in an optical cavity and decoupled from the nanowire [Fig.~\ref{fig:Result1}(a), solid purple line]. This similarity characterizes the single-fermion nature of the state formed from the overlap between the two half-fermionic MZMs.      

The conductance through the QD as a function of both $eV$ and photon occupation, corresponding to the $eV$-range within the gray area in Fig.~\ref{fig:Result2}(a), is shown in Fig.~\ref{fig:Result2}(d).  One can see that the photon-induced transitions in the QD renormalize its spectrum, giving rise to a `bowtie-like' shape, which resembles those found by Prada et al.~\cite{Prada2017}.

In Fig.~\ref{fig:Result3} we display the case in which the MZMs have lost their nonlocal feature ($\eta \rightarrow 1$). This situation qualitatively describes a scenario where the right MZM is displaced towards to left edge of the nanowire, leading to a partial separation between the MZMs and hence a finite overlap between the wavefunction of the right MZM and the QD ($\lambda_{c,R}\gg \varepsilon_M$)~\cite{Prada2017,Penaranda2018,DJClarcke2017}. The emergence of this MZMs with partial separation can result from inhomogeneous confining potentials in the nanowire~\cite{Penaranda2018,prada2019andreev,pan2021quantized,Pan2020}. Similarly to Fig.~\ref{fig:Result2}(a)-(b), we can notice two near-zero-energy conductance peaks at $eV\approx \pm \lambda_{c,R}$ for the case of zero photon occupation [Fig.~\ref{fig:Result3}(a)-(b), dashed red line], which points that the left and right MZMs leak into the dot with different intensities ($\lambda_{c,L}\gg\lambda_{c,R}$) and form a state of a regular fermionic character.

The peak corresponding to the QD conduction level at $eV\approx100\Gamma$ is again splitted into two polariton peaks in presence of cavity photons, which also affects the position of the near-zero-energy conductance peaks [Fig.~\ref{fig:Result3}(b)]. But distinct from the previous case of overlapped MZMs localized at the nanowire edges [Fig.~\ref{fig:Result2}(b)], there is no coalescence into a single peak at $eV = 0$ for certain value of $\langle \nu \rangle$.

Fig.~\ref{fig:Result3}(c) displays the evolution of zero-bias conductance as a function of $\langle \nu \rangle$ for increasing values of coupling between the right-MZM and the QD. For the ideal situation of highly nonlocal MZMs [solid red line], one can see a ZBP plateau associated to the isolated MZM which has leaked into the dot. A plateau is still present for a tiny value of $\lambda_{c,R}$ [dashed orange line], but with a reduction of the ZBP height ($G(eV=0)<0.5e^{2}/h$). However, a little enhancement of $\lambda_{c,R}$ [dotted green and dash-dotted blue lines] drops the zero-bias conductance to almost zero for any value of $\langle \nu \rangle$. 

The behavior of the conductance through the QD as a function of $eV$ and photon occupation $\langle \nu \rangle$ is shown in Fig.~\ref{fig:Result3}(d). One can clearly notice that the near-zero-energy peaks corresponding to these MZMs with less degree of nonlocality slightly move away from each other as photon-induced peak [upper branch in Fig.~\ref{fig:Result3}(d)] is driven towards $eV=0$ and then start to approach each other again as the photon occupation is increased. This behavior yields a pattern that  
resembles the `diamond-like' profile reported by Prada et al.~\cite{Prada2017}. In this work, it was suggested, that the degree of Majorana nonlocality can be extracted from the anticrossing points between the QD level and the near-zero energy levels corresponding to MZMs by application of a QD-gate voltage~\cite{Prada2017,DengPRB2018}. 

Similarly, in the geometry considered by us, one can also extract the information about the degree of Majorana nonlocality from the photon-induced conductance profile of Fig.~\ref{fig:Result3}(d). The dot spectrum now is optically, and not electrically, and the corresponding anticrossing points $\epsilon^{+}_{\text{QD}}$ and $\epsilon^{+}_{\text{MZM}}$ are indicated by dashed green lines. As originally proposed by Prada et al.~\cite{Prada2017}, the degree of Majorana nonlocality in tunneling conductance profiles~\cite{DengPRB2018} is given by $\Omega_{M}^{2}\approx\eta^{2}=\epsilon^{+}_{\text{MZM}}/\epsilon^{+}_{\text{QD}}$, valid when $\varepsilon_{M}\ll \lambda_{c,L}, \lambda_{c,R}$. For the parameters corresponding to Fig.~\ref{fig:Result3}(d), $\Omega_{M}\approx 0.35$, or equivalently, a topological quality factor of $\approx 0.88$~\cite{DJClarcke2017,RiccoPRB2020}, thus indicating that the MZMs are not well-localized at the nanowire edges. 

Once we cannot extract the information about the wavefunctions of the Majorana nanowire within the effective model employed here, we are not able to do a detailed numerical analysis of the degree of Majorana nolocality as done by Prada et al~\cite{Prada2017}. However, in Sec.~\ref{AppB} of the Appendix we performed a theoretical analysis of the effective spinless Hamiltonian considered throughout this work, showing the matching between the anticrossing points $\epsilon_{\text{QD,MZM}}^{\pm}$ analytically obtained to extract $\Omega_{M}$ within our proposal and those ones originally proposed in Ref.~\onlinecite{Prada2017}. 

\textbf{Conclusions}

In summary, we have theoretically explored the effects of strong light-matter coupling on transport properties of the hybrid device, consisting on quantum dot embedded inside a single-mode optical cavity and coupled to a Majorana nanowire. Conductance profiles as functions of the bias-voltage and mean photon occupation number controlled by an external pump revealed distinct shapes for the cases of highly nonlocal MZMs, overlapped MZMs localized at opposite nanowire edges and MZMs which are not perfectly nonlocal. This makes possible to access the degree of Majorana nonlocality (topological quality factor) all optically, by means of the tuning of the polariton energies in the dot.

\textbf{Methods}

The full Hamiltonian which describes the system of Fig.~\ref{fig:System} reads ($\hbar = 1$):
\begin{equation}
\mathcal{H}=\mathcal{H}_{ph}+\mathcal{H}_{QD}+\mathcal{H}_{int}+\mathcal{H}_{lead}+\mathcal{H}_{MZMs},\label{eq:H_full}
\end{equation}
where $\mathcal{H}_{ph}=\omega_{0}c^{\dagger}c,$ is the Hamiltonian of a single-mode optical cavity, with  $c^{\dagger},(c)$ being creation and annihilation operators of the cavity photons with energy $\omega_{0}$. The Hamiltonian of the QD reads 
$\mathcal{H}_{QD}=\omega_{c}d_{c}^{\dagger}d_{c}+\omega_{v}d_{v}^{\dagger}d_{v}$, 
where $d_{v}$ and $d_{c}$ describe the spinless electrons in the valence and conduction levels of the QD with energy $\omega_c$ and $\omega_v$, respectively, and $\Omega = \omega_c-\omega_v$ is the energy difference between these levels. The spinless situation can be reached in the regime of large magnetic fields, where only either the spin up or down channel is accounted. In this regime, only single occupancy is allowed at each QD level and therefore the onsite Coulomb repulsion can be safely neglected in $\mathcal{H}_{QD}$. Otherwise, the assumption of both spin degrees of freedom with Coulomb interaction between electrons at each QD level lead to the appearance of extra peaks in the QD energy spectrum, known as Hubbard peaks~\cite{RiccoPRB2019,TijerinaPRB2015,Hubbard1963}. Moreover, the presence of such a Coulomb repulsion also can lead to Kondo-type correlations for $T< T_{K}$~\cite{CronenwettScience1998,Goldhaber1998} or $T_{K}/\Delta_{Nw}\gtrsim 0.6$~\cite{Prada2017,LeePRB}, where $T$ is the system temperature, $T_{K}$ is the characteristic Kondo temperature and $\Delta_{Nw}$ is the induced superconducting pairing in the Majorana nanowire.

The interaction between the cavity and the QD is given by~\cite{Bruce1993}   
\begin{equation}
\mathcal{H}_{int}=-\Omega_{R}(d_{v}^{\dagger}d_{c}c^{\dagger}+d_{c}^{\dagger}d_{v}c),\label{H_int}
\end{equation}
where $\Omega_{R}$ is the Rabi splitting strength.

The source (S) and drain (D) metallic leads and their coupling with the QD are described by
\begin{equation}
\mathcal{H}_{lead}=\sum_{\boldsymbol{k},\alpha}\varepsilon_{\boldsymbol{k}}^{\alpha}c_{\boldsymbol{k},\alpha}^{\dagger}c_{\boldsymbol{k},\alpha}+\sum_{\boldsymbol{k},\alpha,l}V_{l}(c_{\boldsymbol{k},\alpha}^{\dagger}d_{l}+\text{h.c}),\label{H_lead}
\end{equation}
where $c_{\boldsymbol{k},\alpha}^{\dagger}$ ($c_{\boldsymbol{k},\alpha}$) creates (annihilates) an electron in the lead $\alpha=S/D$, with wave-number $\boldsymbol{k}$, energy $\varepsilon_{\boldsymbol{k}}^{\alpha} = \epsilon_{\boldsymbol{k}}-\mu_{\alpha}$ and chemical potential $\mu_{\alpha}$. The bias-voltage through the QD is defined as $eV=\mu_{S}-\mu_{D}$. The parameter $V_{l}$ represents the coupling strength of the conduction (${l=c}$) and valence (${l=v}$) levels of the QD with the leads. Once we are considering the situation in which the valence level is far below the Fermi level $(\omega_{v}\ll E_{F}, E_{F} = 0)$ [Fig.~\ref{fig:System}(b)], one can assume $V_v=0$ and $V_c = \mathcal{V}$ in Eq.~(\ref{H_lead}) without loss of generality. Introducing even and odd linear combinations of the states of the leads $c_{\boldsymbol{k},e}$,$c_{\boldsymbol{k},o}$, i.e. performing the unitary transformation according to $c_{\boldsymbol{k},S}=\left(c_{\boldsymbol{k},o}+c_{\boldsymbol{k},e}\right)/\sqrt{2}$ and $c_{\boldsymbol{k},D}=\left(c_{\boldsymbol{k},e}-c_{\boldsymbol{k},o}\right)/\sqrt{2},$ the odd states become decoupled from the dot, and the Hamiltonian of Eq.~(\ref{H_lead}) transforms into~\cite{RiccoOscillations2018}:
\begin{equation}
\mathcal{H}_{lead}=\sum_{\boldsymbol{k},a=e,o}\epsilon_{\boldsymbol{k}}c_{\boldsymbol{k},a}^{\dagger}c_{\boldsymbol{k},a}+\sqrt{2}\mathcal{V}\sum_{\boldsymbol{k}}(c_{\boldsymbol{k},e}^{\dagger}d_{c}+\text{h.c}).\label{eq:H_lead_even}
\end{equation}

The effective Hamiltonian which describes the `half-fermionic' states corresponding to MZMs at the ends of the Majorana nanowire coupled to the QD reads~\cite{Prada2017,DJClarcke2017}:
\begin{eqnarray}
\mathcal{H}_{MZMs}& =& \imath\varepsilon_{M}\gamma_{L}\gamma_{R}+\lambda_{c,L}(d_{c}-d_{c}^{\dagger})\gamma_{L}\nonumber \\ 
 & +& \imath \lambda_{c,R}(d_c + d_{c}^{\dagger})\gamma_{R}, \label{eq:H_MBS}
\end{eqnarray}
where $\gamma_{L,R}=\gamma_{L,R}^\dagger$ represent the MZMs at the opposite ends of the Majorana nanowire with $\varepsilon_{M}$ being the overlap strength between them~\cite{RevMajoranaAguado}. The hybridization between the left and right MZMs with the conduction level of the dot is given by $\lambda_{c,L}$ and $\lambda_{c,R}$, respectively, and one can safely assume, that $\lambda_{v,L(R)}=0$ for the situation of $\omega_{v}\ll E_{F}$ considered by us, once the MZMs cannot overlap with QD valence level owing to the large energy separation between them. Moreover, the assumption of $\omega_{v}$ far below the Fermi level and consequent zero coupling strength between the MZMs and the QD valence level ensures the resonance condition $\Omega=\omega_{c}-\omega_{v}=\omega_{0}$ within a range of realistic parameters, i.e., $\Omega=\omega_{0}\sim\SI{}{\electronvolt}$~\cite{Khitrova2006}. 

The Hamiltonian of Eq.~(\ref{eq:H_MBS}) can be rewritten in terms of a fermionic operator $f$~\cite{RevMajoranaAlicea,RevMajoranaAguado} by considering the fermionic representation of Majorana operators $\gamma_{L} =(f^{\dagger}+f)/\sqrt{2}$ and $\gamma_{R} =\imath(f^{\dagger}-f)/\sqrt{2}$, where $f$ obeys the standard fermionic anticommutation relations. Hence, Eq.~(\ref{eq:H_MBS}) becomes into $
\mathcal{H}_{MZMs}=\varepsilon_{M}f^{\dagger}f + (t_{c}d_{c}f^{\dagger}+\Delta_{c}d_{c}f+\text{h.c.}),$
with $t_{c}=(\lambda_L - \lambda_R)/\sqrt{2}$ and $\Delta_{c}=(\lambda_L + \lambda_R)/\sqrt{2}$.

The application of a bias-voltage $eV$ between source and drain leads, leads to the onset of the current through the system, and, according to the Landauer-type formula~\cite{MeirPRL1992,Bruus}, at low temperatures ($T\rightarrow 0$) the conductance through the QD reads:
\begin{equation}
G(eV)=\left(\frac{e^{2}}{h}\right)\pi\Gamma\rho_{c}(eV)\label{eq:Conductance_T=0_only_Ec}
\end{equation}
%
%
where $e^{2}/h$ is the quantum of conductance and $\Gamma=2\pi \mathcal{V}^{2}\rho$ represents the effective broadening introduced by the coupling between the QD and the even conduction operator of the leads [Eq.~(\ref{eq:H_lead_even})], with a constant density of states $\rho$~\cite{Anderson,Bruus}, valid within the the wide-band limit approximation. The local density of states of the dot is given by~\cite{Bruus}: 
\begin{equation}
\rho_{i}(\omega)=-\frac{1}{\pi}\text{Im}\langle\langle d_{c};d_{c}^{\dagger}\rangle\rangle_{\omega}, \label{eq:LDOS}
\end{equation}
where $\langle\langle d_{c};d_{c}^{\dagger}\rangle\rangle_{\omega}\equiv\boldsymbol{G}^{c}(\omega)$ is the retarded Green's function in the spectral domain $\omega$~\cite{Zubarev1960}, which can be calculated by successive applications of equation-of-motion technique~\cite{Bruus,Zubarev1960,Jauho} together with an appropriate truncation scheme (see Sec.~\ref{AppA} of Appendix for details concerning the equation of motion technique and the truncation scheme adopted). This gives
\begin{equation}
\boldsymbol{G}^{c}(\omega)=\frac{\boldsymbol{g}_{0}(\omega)}{1+\boldsymbol{g}_{0}(\omega)\left[\imath\Gamma-\Sigma_{ph}^{c}(\omega)-\Sigma_{MZMs}(\omega)\right]},
\label{eq:GF_conduction}
\end{equation}
where $\boldsymbol{g}_{0}(\omega) = 1/(\omega + \imath \delta - \omega_{c})$ is the \textit{bare} Green's function of the QD conduction level~\cite{Bruus,Dyson1949,Odashima2017}, with $\delta \rightarrow 0^{+}$. The term $\imath\Gamma$ describes the corresponding broadening introduced by the coupling with metallic leads and
\begin{equation}
\Sigma_{ph}^{c}(\omega)=\frac{\Omega_{R}^{2}\langle\nu\rangle}{(\omega+i\delta-\omega_{0}-\omega_{v})}\label{eq:Sigmaphc}
\end{equation}
is the self-energy associated to the valence-to-conduction band transition induced by the photonic field of the optical cavity, with $\langle \nu \rangle = \langle n_{c} \rangle + N_{ph} $ being the mean photon occupation, which depends on the number of excitations ($\langle n_{c} \rangle$) in the QD and photons ($N_{ph}$) in the microcavity, which can be tunned by an external optical pump. 

The part of the self-energy associated with the direct coupling between the QD conduction level and MZMs reads:
\begin{equation}
\Sigma_{MZMs}(\omega)=\kappa_{1}(\omega) + (t_c\Delta_c)^{2}\kappa_{0}(\omega)K(\omega),\label{eq:SigmaMBSs} 
\end{equation}
where $\kappa_{0}(\omega)=[(\omega+i\delta+\varepsilon_{M})^{-1}+(\omega+i\delta-\varepsilon_{M})^{-1}]$, $\kappa_{1}(\omega)=[t_{c}^{2}(\omega+i\delta-\varepsilon_{M})^{-1}+\Delta_{c}^{2}(\omega+i\delta+\varepsilon_{M})^{-1}]$, $\tilde{\kappa}_{1}(\omega)=[\Delta_{c}^{2}(\omega+i\delta-\varepsilon_{M})^{-1}+t_{c}^{2}(\omega+i\delta+\varepsilon_{M})^{-1}]$, $K(\omega) = \kappa_{0}(\omega)/(\omega + \imath \Gamma + \omega_{c} - \tilde{\Sigma}_{ph}^{c}(\omega) - \tilde{\kappa}_{1}(\omega))$, with $\tilde{\Sigma}_{ph}^{c}(\omega) = \Omega_{R}^{2}\langle \nu \rangle/(\omega + \imath\delta + \omega_{0} + \omega_{v})$ . If the QD is decoupled from the optical cavity ($\Omega_{R} = 0$), $\tilde{\Sigma}_{ph}^{c}(\omega) = 0$ and  Eq.~(\ref{eq:SigmaMBSs}) is reduced to the well-known expression for the self-energy associated to the leaking of a single MZM into the QD for $\lambda_{c,R}=0$~\cite{VernekPRB2014,DengScience2016,Baranger2011,VLCampoPRB2017}. 

The presence of $\tilde{\Sigma}_{ph}^{c}(\omega)$ in the expression for $K(\omega)$ in the self-energy defined in Eq.~(\ref{eq:SigmaMBSs}) means that the MZMs somehow `feel' the photonic field, although there is no direct coupling between cavity photons and the Majorana nanowire.

\medskip
\textbf{Data Availability}

The data that support the findings of this study are available from the corresponding author upon reasonable request.

\medskip
\textbf{Acknowledgments}

LSR, VKK and IAS acknowledge support from Icelandic Research Fund (project ``Hybrid polaritonics'') and Russian Science Foundation (project 20-12-00224). ACS acknowledges support from Brazilian National Council for Scientific and Technological Development (CNPq), grant~305668/2018-8.

\appendix

\section{Conduction band Green's function derivation}\label{AppA}

The quantum dot conduction band Green's function $\boldsymbol{G}^{c}(\omega)\equiv\langle\langle d_{c};d_{c}^{\dagger}\rangle\rangle_{\omega}$ [Eq.~(8) in the main text] can be derived through successive applications of the equation-of-motion (EOM) technique~\cite{Bruus,Zubarev1960}. For retarded Green's functions in the spectral domain, the EOM reads
\begin{equation}
	(\omega+\imath\delta)\boldsymbol{G}_{A_{i},B_{j}}(\omega)=[A_{i},B_{j}^{\dagger}]_{+}+\langle\langle[A_{i},\mathcal{H}];B_{j}^{\dagger}\rangle\rangle_{\omega},\label{eq:EOM}
\end{equation}
where $\boldsymbol{G}_{A_{i},B_{j}}(\omega)\equiv\langle\langle A_{i};B_{j}^{\dagger}\rangle\rangle_{\omega}$ is the retarded Green's function in the notation adapted from Zubarev~\cite{Zubarev1960,Odashima2017}, $\delta \rightarrow 0^{+}$ is a positive infinitesimal number, $A_{i}$ and $B_{j}$ are operators belonging to the Hamiltonian $\mathcal{H}$ of the system [Eq.~(1) in the main text] and $[\cdots,\cdots]_{+}$ is the standard anticommutation relation for fermions~\cite{Bruus}. Considering $A_{i}=B_{j}=d_{c}$:
\begin{eqnarray}
	(\omega-\omega_{c}+\imath\delta)\boldsymbol{G}^{c}(\omega) & = & 1+\sqrt{2}\mathcal{V}\sum_{\boldsymbol{k}}\langle\langle c_{\boldsymbol{k},e};d_{c}^{\dagger}\rangle\rangle\nonumber\\
	&-&\Omega_{R}\langle\langle d_{v}c;d_{c}^{\dagger}\rangle\rangle_{\omega}\nonumber\\
	&-&t_{c}\langle\langle f;d_{c}^{\dagger}\rangle\rangle_{\omega}-\Delta_{c}\langle\langle f^{\dagger};d_{c}^{\dagger}\rangle\rangle_{\omega}, \nonumber\\
	& &\label{eq:EOM1}
\end{eqnarray}
with $t_{c}=(\lambda_L - \lambda_R)/\sqrt{2}$, $\Delta_{c}=(\lambda_L + \lambda_R)/\sqrt{2}$ and $f = (\gamma_{L} + \imath \gamma_{R})/\sqrt{2} $ is a complex fermionic operator built from combination of left ($\gamma_{L}$) and right ($\gamma_{R}$) MZMs~\cite{RevMajoranaAguado}, and can acquire a nonlocal feature depending on the distance between these `half-fermionic' Majorana states. As noticed in Eq.~(\ref{eq:EOM1}), the high-order Green's function $\langle\langle d_{v}c;d_{c}^{\dagger}\rangle\rangle_{\omega}$ arises due to the light-matter coupling term given by $\mathcal{H}_{int}$ [Eq.~(2) in the main text]. According to the EOM:
\begin{widetext}
\begin{equation}
	(\omega+\imath\delta)\langle\langle d_{v}c;d_{c}^{\dagger}\rangle\rangle_{\omega}=[d_{v}c,d_{c}^{\dagger}]_{+}+\langle\langle[d_{v}c,\mathcal{H}];d_{c}^{\dagger}\rangle\rangle_{\omega}=\langle\langle[d_{v}c,\mathcal{H}];d_{c}^{\dagger}\rangle\rangle_{\omega}\label{eq:EOM2},
\end{equation}
once $[d_{v}c,d_{c}^{\dagger}]_{+}=0$ and the commutation relation
\begin{equation}
	[d_{v}c,\mathcal{H}]=\omega_{0}d_{v}c+\omega_{v}d_{v}c-\Omega_{R}(d_{v}d_{v}^{\dagger}d_{c}+c^{\dagger}cd_{c}).
\end{equation}
Thus, Eq.~(\ref{eq:EOM2}) becomes
\begin{equation}
	(\omega-\omega_{0}-\omega_{v}+\imath\delta)\langle\langle d_{v}c;d_{c}^{\dagger}\rangle\rangle_{\omega}=-\Omega_{R}\langle\langle d_{v}d_{v}^{\dagger}d_{c};d_{c}^{\dagger}\rangle\rangle_{\omega}-\Omega_{R}\langle\langle c^{\dagger}cd_{c};d_{c}^{\dagger}\rangle\rangle_{\omega},\label{eq:EOM3}
\end{equation}
where new high-order Green's functions $\langle\langle d_{v}d_{v}^{\dagger}d_{c};d_{c}^{\dagger}\rangle\rangle_{\omega}$ and $\langle\langle c^{\dagger}cd_{c};d_{c}^{\dagger}\rangle\rangle_{\omega}$ arises. In order to find an analytical expression for $\langle\langle d_{v}c;d_{c}^{\dagger}\rangle\rangle_{\omega}$, we prevent the emergence of more high-order Green's functions in the EOM process by considering the following truncation:
\begin{eqnarray}
	\langle\langle d_{v}d_{v}^{\dagger}d_{c};d_{c}^{\dagger}\rangle\rangle_{\omega}& = &\langle d_{v}d_{v}^{\dagger}\rangle\langle\langle d_{c};d_{c}^{\dagger}\rangle\rangle_{\omega}=(1-\langle d_{v}^{\dagger}d_{v}\rangle)\langle\langle d_{c};d_{c}^{\dagger}\rangle\rangle_{\omega}\nonumber\\
	&=&\langle d_{c}^{\dagger}d_{c}\rangle\langle\langle d_{c};d_{c}^{\dagger}\rangle\rangle_{\omega},\label{truncation1}
\end{eqnarray}
wherein we considered the one-electron assumption $\langle d_{c}^{\dagger}d_{c}\rangle+\langle d_{v}^{\dagger}d_{v}\rangle=1$ valid for all times~\cite{Bruce1993}
and 
\begin{equation}
	\langle\langle c^{\dagger}cd_{c};d_{c}^{\dagger}\rangle\rangle_{\omega}=\langle c^{\dagger}c\rangle\langle\langle d_{c};d_{c}^{\dagger}\rangle\rangle_{\omega}\label{truncation2}.
\end{equation}
Without loss of generality, in the truncation scheme of Eqs.~(\ref{truncation1}) and~(\ref{truncation2}) we turned the high-order Green's functions of Eq.~(\ref{eq:EOM3}) into the conduction level Green's function $\langle\langle d_{c};d_{c}^{\dagger}\rangle\rangle_{\omega}$ modulated by the associated mean number of excitations in the QD $\langle d_{c}^{\dagger}d_{c}\rangle=\langle n_c \rangle$ and the mean number of photons within the optical cavity $\langle c^{\dagger}c\rangle = N_{ph}$, respectively. Thus, Eq.~(\ref{eq:EOM3}) reads
\begin{equation}
	(\omega-\omega_{0}-\omega_{v}+\imath\delta)\langle\langle d_{v}c;d_{c}^{\dagger}\rangle\rangle_{\omega}=-\Omega_{R}(\langle n_{c}\rangle+N_{ph})\langle\langle d_{c};d_{c}^{\dagger}\rangle\rangle_{\omega}.\label{EOM4}
\end{equation}
Considering that the $\langle n_{c} \rangle + N_{ph}=\langle\nu\rangle$ gives the mean photon occupation in the cavity, Eq.~(\ref{EOM4}) turns into 
\begin{equation}
	\langle\langle d_{v}c;d_{c}^{\dagger}\rangle\rangle_{\omega}=\frac{-\Omega_{R}\langle\nu\rangle}{(\omega-\omega_{0}-\omega_{v}+\imath\delta)}\langle\langle d_{c};d_{c}^{\dagger}\rangle\rangle_{\omega}.\label{EOM5}
\end{equation}
Substituting Eq.~(\ref{EOM5}) into (\ref{eq:EOM1}), we find
\begin{eqnarray}
	(\omega-\omega_{c}+\imath\delta)\boldsymbol{G}^{c}(\omega)  & = &  1+\sqrt{2}\mathcal{V}\sum_{\boldsymbol{k}}\langle\langle c_{\boldsymbol{k},e};d_{c}^{\dagger}\rangle\rangle+\Sigma_{ph}^{c}(\omega)\boldsymbol{G}^{c}(\omega)\nonumber\\&-&t_{c}\langle\langle f;d_{c}^{\dagger}\rangle\rangle_{\omega}-\Delta_{c}\langle\langle f^{\dagger};d_{c}^{\dagger}\rangle\rangle_{\omega} \label{EOM6},
\end{eqnarray}
where $\Sigma_{ph}^{c}(\omega)=\frac{\Omega_{R}^{2}\langle\nu\rangle}{\omega-\omega_{0}-\omega_{v}+i\delta}$ [Eq.~(9) in the main text] is the self-energy associated to the valence-to-conduction level transition in the quantum dot induced by the single-mode photonic field of the cavity.

At this point, the Green's function of Eq.~(\ref{EOM6}) that mixes the even conduction operator $c_{\boldsymbol{k},e}$ from the leads with the operator $d_{c}$ from the dot should be calculated. According to the EOM procedure, this Green's function reads
\begin{equation}
	\langle\langle c_{\boldsymbol{k},e};d_{c}^{\dagger}\rangle\rangle_{\omega}=\frac{\sqrt{2}\mathcal{V}}{\omega+\imath\eta-\epsilon_{\boldsymbol{k}}}\langle\langle d_{c};d_{c}^{\dagger}\rangle\rangle_{\omega}\label{eq:ck,dc}
\end{equation}
Substitution of the expression above into Eq.~(\ref{EOM6}) yields
\begin{eqnarray}
	(\omega-\omega_{c}+\imath\delta)\boldsymbol{G}^{c}(\omega) & = & 1+\Sigma_{lead}(\omega)\boldsymbol{G}^{c}(\omega)+\Sigma_{ph}^{c}(\omega)\boldsymbol{G}^{c}(\omega)\nonumber\\
	&-&t_{c}\langle\langle f;d_{c}^{\dagger}\rangle\rangle_{\omega}-\Delta_{c}\langle\langle f^{\dagger};d_{c}^{\dagger}\rangle\rangle_{\omega} \label{11},
\end{eqnarray}
where $\Sigma_{\text{lead}}(\omega)=\sum_{\boldsymbol{k}}\frac{2\mathcal{V}^{2}}{\omega+\imath\eta-\epsilon_{\boldsymbol{k}}}$ is the self-energy due to the coupling between the QD and the even conduction operators of metallic leads [see Eq.(4) of main text]. In the wide-band limit, $\text{Re}[\Sigma_{\text{lead}}(\omega)]\rightarrow 0$ and this self-energy is reduced to $-i\Gamma$, which is independent of $\omega$, with $\Gamma = 2\pi \mathcal{V}^{2}\rho$~\cite{Anderson} as defined in the main text. Thus, Eq.~(\ref{11}) reads
\begin{eqnarray}
	(\omega-\omega_{c}+\imath\delta)\boldsymbol{G}^{c}(\omega)  & = & 1-\imath \Gamma \boldsymbol{G}^{c}(\omega)+\Sigma_{ph}^{c}(\omega)\boldsymbol{G}^{c}(\omega)-t_{c}\langle\langle f;d_{c}^{\dagger}\rangle\rangle_{\omega}\nonumber\\
	&-&\Delta_{c}\langle\langle f^{\dagger};d_{c}^{\dagger}\rangle\rangle_{\omega} \label{13}
\end{eqnarray}

Also through application of EOM [Eq.~(\ref{eq:EOM})], the Green's functions of Eq.~(\ref{EOM6}) which mix the operator of the dot conduction level with the fermionic operator built from Majorana modes are given by
\begin{equation}
	\langle\langle f;d_{c}^{\dagger}\rangle\rangle_{\omega}=\frac{-t_{c}\langle\langle d_{c};d_{c}^{\dagger}\rangle\rangle_{\omega}+\Delta_{c}\langle\langle d_{c}^{\dagger};d_{c}^{\dagger}\rangle\rangle_{\omega}}{\omega-\varepsilon_{M}+\imath\delta}\label{EOM7}
\end{equation}
and 
\begin{equation}
	\langle\langle f^{\dagger};d_{c}^{\dagger}\rangle\rangle_{\omega}=\frac{t_{c}\langle\langle d_{c}^{\dagger};d_{c}^{\dagger}\rangle\rangle_{\omega}-\Delta_{c}\langle\langle d_{c};d_{c}^{\dagger}\rangle\rangle_{\omega}}{\omega+\varepsilon_{M}+\imath\delta}.\label{EOM8}
\end{equation}
Hence, Eq.~(\ref{13}) reads
\begin{eqnarray}
	(\omega-\omega{}_{c}+i\delta)\boldsymbol{G}^{c}(\omega)&=&1-\imath\Gamma\boldsymbol{G}^{c}(\omega)+\Sigma_{ph}^{c}(\omega)\boldsymbol{G}^{c}(\omega)+\kappa_{1}(\omega)\boldsymbol{G}^{c}(\omega)\nonumber\\
	&-&t_{c}\Delta_{c}\kappa_{0}(\omega)\langle\langle d_{c}^{\dagger};d_{c}^{\dagger}\rangle\rangle_{\omega},\label{EOM9}
\end{eqnarray}
with
\begin{equation}
	\kappa_{1}(\omega)=\frac{t_{c}^{2}}{\omega-\varepsilon_{M}+\imath\delta}+\frac{\Delta_{c}^{2}}{\omega+\varepsilon_{M}+\imath\delta}
\end{equation}
and
\begin{equation}
 \kappa_{0}(\omega)=\frac{1}{\omega-\varepsilon_{M}+\imath\delta}+\frac{1}{\omega+\varepsilon_{M}+\imath\delta}.\label{EOM10}   
\end{equation}
In Eq.~(\ref{EOM9}), one can notice the presence of the retarded Green's function $\langle\langle d_{c}^{\dagger};d_{c}^{\dagger}\rangle\rangle_{\omega}$ associated to the superconductivity in the quantum dot induced by the coupling with the Majorana nanowire ~\cite{Baranger2011,VernekPRB2014,RiccoPRB2019,RiccoOscillations2018} and, according to the EOM technique, is given by
\begin{eqnarray}
	(\omega+\omega_{c}+\imath\delta)\langle\langle d_{c}^{\dagger};d_{c}^{\dagger}\rangle\rangle_{\omega} &= & -\imath\Gamma \langle\langle d_{c}^{\dagger};d_{c}^{\dagger}\rangle\rangle_{\omega} + \tilde{\Sigma}_{ph}^{c}(\omega)\langle\langle d_{c}^{\dagger};d_{c}^{\dagger}\rangle\rangle_{\omega}\nonumber\\
	& +& t_{c}\langle\langle f^{\dagger};d_{c}^{\dagger}\rangle\rangle_{\omega}
	 +\Delta_{c}\langle\langle f;d_{c}^{\dagger}\rangle\rangle_{\omega},\label{EOM11}
\end{eqnarray}
where $\tilde{\Sigma}_{ph}^{c}(\omega)=\frac{\Omega_{R}^{2}\langle\nu\rangle}{\omega+\omega_{0}+\omega_{v}+\imath\delta}$ as defined in the main text. Substituting Eqs.~(\ref{EOM7}) and~(\ref{EOM8}) into Eq.~(\ref{EOM11}), we find
\begin{equation}
	\langle\langle d_{c}^{\dagger};d_{c}^{\dagger}\rangle\rangle_{\omega}=-t_{c}\Delta_{c}K(\omega)\boldsymbol{G}^{c}(\omega),\label{EOM12}
\end{equation}
with
\begin{equation}
	K(\omega)=\frac{\kappa_{0}(\omega)}{\omega+\imath\Gamma +\omega_{c}-\tilde{\Sigma}_{ph}^{c}(\omega)-\tilde{\kappa}_{1}(\omega)}\label{K}
\end{equation}
and
\begin{equation}
 \tilde{\kappa}_{1}(\omega)=\frac{t_{c}^{2}}{\omega+\varepsilon_{M}+\imath\delta}+\frac{\Delta_{c}^{2}}{\omega-\varepsilon_{M}+\imath\delta}.\label{K_1tilde}   
\end{equation}
Straightforward substitution of Eq.~(\ref{EOM12}) into Eq.~(\ref{EOM9}) yields
\begin{equation}
	\boldsymbol{G}^{c}(\omega)= \boldsymbol{g}_0(\omega) + \boldsymbol{g}_0(\omega)\left[ \Sigma_{ph}^{c}(\omega)+\Sigma_{MZMs}(\omega) -\imath \Gamma \right]\boldsymbol{G}^{c}(\omega),
	\label{eq:GF_conduction}
\end{equation}
as the Dyson equation~\cite{Bruus,Dyson1949} for the quantum dot conduction band, with the self-energy [Eq.~(10) in the main text] \begin{equation}
	\Sigma_{MZMs}(\omega)=\kappa_{1}(\omega) + (t_c\Delta_c)^{2}\kappa_{0}(\omega)K(\omega)\label{MZMs_self}    
\end{equation}
responsible for renormalizing the dot energy spectrum due to the coupling with the MZMs located at the nanowire. The presence of $\tilde{\Sigma}_{ph}^{c}(\omega)$ within $K(\omega)$ [Eq.~(\ref{K})] in the resulting self-energy of Eq.~(\ref{MZMs_self}) reveals that although the Majorana nanowire is decoupled from the optical cavity, the MZMs are indirectly affected by the photon-induced transitions in the quantum dot.

Finally, by isolating $\boldsymbol{G}^{c}(\omega)$ in Eq.~(\ref{eq:GF_conduction}), we can write the expression for the Green's function of the quantum dot conduction band as 
\begin{equation}
	\boldsymbol{G}^{c}(\omega)=\frac{\boldsymbol{g}_{0}(\omega)}{1+\boldsymbol{g}_{0}(\omega)\left[\imath\Gamma-\Sigma_{ph}^{c}(\omega)-\Sigma_{MZMs}(\omega)\right]},
\end{equation}
which is the same relation of Eq.~(8) in the main text.
\end{widetext}

\section{Analysis of anticrossing points related to the degree of Majorana nonlocality}\label{AppB}

Following the same procedure of Prada et al.~\cite{Prada2017}, the effective Hamiltonian which describes the QD conduction level coupled to both the left and right MZMs of the Majorana nanowire reads
\begin{equation}
\mathcal{H}^{\text{eff}}=\omega_{c}d_{c}^{\dagger}d_{c}+\imath\varepsilon_{M}\gamma_{L}\gamma_{R}+\lambda_{c,L}(d_{c}-d_{c}^{\dagger})\gamma_{L}+\imath\lambda_{c,R}(d_{c}+d_{c}^{\dagger})\gamma_{R}.
\end{equation}

This Hamiltonian can be rewritten as

\begin{equation}
\mathcal{H}^{\text{eff}}=\frac{1}{2}\psi^{\dagger}\check{\mathcal{H}}^{\text{eff}}\psi,
\end{equation}
with $\psi=\begin{pmatrix}d_{c} & d_{c}^{\dagger} & \gamma_{L} & \gamma_{R}\end{pmatrix}^{T}$ and

\begin{equation}
\check{\mathcal{H}}^{\text{eff}}=\begin{pmatrix}\omega_{c} & 0 & -\lambda_{c,L} & \imath\lambda_{c,R}\\
0 & -\omega_{c} & \lambda_{c,L} & \imath\lambda_{c,R}\\
-\lambda_{c,L} & \lambda_{c,L} & 0 & \imath\varepsilon_{M}\\
-\imath\lambda_{c,R} & -\imath\lambda_{c,R} & -\imath\varepsilon_{M} & 0 
\end{pmatrix}.\label{eq:HeffMatrix}
\end{equation}

\begin{figure}[t]
	\centerline{\includegraphics[width=3.6in,keepaspectratio]{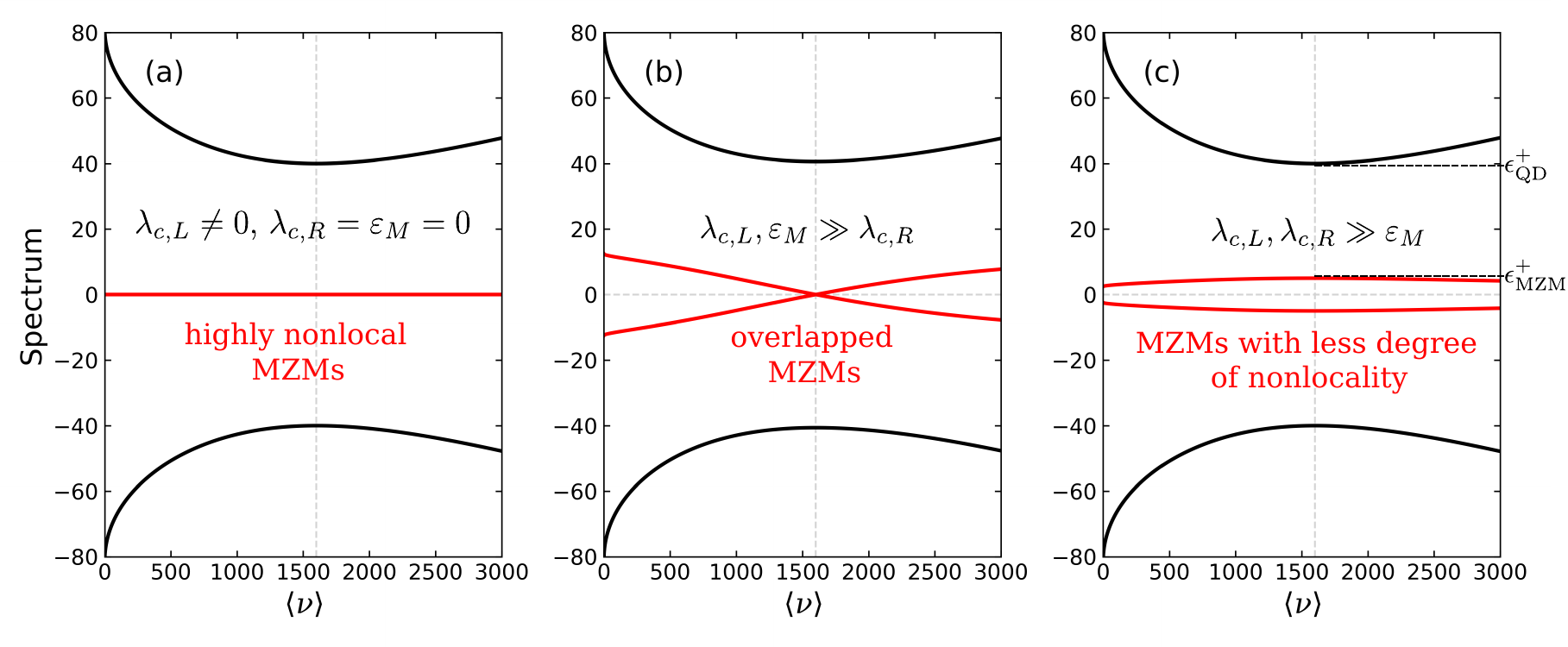}}
	\caption{Low-energy spectrum of the system as a function of mean number of photons $\langle \nu \rangle$, provided by eigenvalues of Eq.~(\ref{eq:HeffMatrix}). The parameters for panels (a), (b) and (c) are the same adopted in Figs.~2(d), 3(d) and 4(d) of the main text, respectively. The gray dashed lines of each panel depicts the points of crossing/anticrossing at $\langle \nu \rangle = (\omega_{c}/\Omega_{R})^{2}$. For $\omega_{c}=100\Gamma$ and $\Omega_{R}=2.5\Gamma$, $\langle \nu \rangle = 1600$. The anticrossing points $\epsilon_{\text{QD,MZM}}^{+}$ of panel (c) are given by the analytical expressions of Eqs.~(\ref{eq:epsilonQD}) and (\ref{eq:epsilonMZM}).}
	\label{fig:SIAnalyticalSpectrum}
\end{figure}

It can be easily noticed that the mean photon occupation $\langle \nu \rangle$ in the microcavity plays no role in Eq.~(\ref{eq:HeffMatrix}). However, it can be seen in Figs.~3(d) and 4(d) of the main text that the crossing/anticrossing points are localized at $\langle \nu \rangle =(\omega_{c}/\Omega_{R})^{2}$. Thus, the main effect of the cavity photons shown in the gray region of Figs.~2(a), 3(a) and 4(a) can be accounted in $\check{\mathcal{H}}^{\text{eff}}$ by considering $\omega_{c}\rightarrow \tilde{\omega}_{c} = \omega_{c} - \Omega_{R}\sqrt{\langle \nu \rangle}$. Although the rightmost peaks of Figs.~2(a), 3(a) and 4(a) are not described by this renormalization, the low-energy spectrum of the system related to the degree of Majorana nonlocality is given by the eigenvalues of the above matrix, which reads
\begin{widetext}
\begin{equation}
E_{\pm(\pm)}=\pm\frac{1}{\sqrt{2}}\left[\sqrt{\left(\frac{(\tilde{\omega}_{c}^{2}+\varepsilon_{M}^{2})}{2}+\lambda_{L}^{2}+\lambda_{R}^{2}\right)(\pm)\sqrt{\left(\frac{\tilde{\omega}_{c}^{2}+\varepsilon_{M}^{2}}{2}+\lambda_{L}^{2}+\lambda_{R}^{2}\right)^{2}-4\left(\frac{\varepsilon_{M}\tilde{\omega}_{c}}{2}+\lambda_{L}\lambda_{R}\right)^{2}}}\right].\label{eq:Eigenvalues}
\end{equation}
\end{widetext}

Fig.~\ref{fig:SIAnalyticalSpectrum} shows exactly the low-energy spectrum given by Eq.~(\ref{eq:Eigenvalues}) as a function of the mean number of photons $\langle \nu \rangle$ in the cavity, for the same situations explored in the main text. One can see a perfect matching between Figs.~\ref{fig:SIAnalyticalSpectrum} (a), (b) and (c) and Figs. 2(d), 3(d) and 4(d) of the main text, respectively, thus revealing that $\check{\mathcal{H}}^{\text{eff}}$ with the QD conduction level renormalized by the cavity photons $\tilde{\omega}_{c}$ indeed describes the main behavior of the conductance of the system obtained via Green's functions 

As stated above, the anticrossing points are localized at $\langle \nu \rangle = (\omega_{c}/\Omega_{R})^{2}$, i.e, for $\tilde{\omega}_{c}=0$. By applying this condition in the corresponding eigenvalues of the effective Hamiltonian [Eq.~(\ref{eq:Eigenvalues})], we are able to find the following general analytical expressions for $\epsilon_{\text{QD,MZM}}^{\pm}$:
\begin{widetext}
\begin{equation}
\epsilon_{\text{QD}}^{\pm}=\pm\frac{1}{\sqrt{2}}\sqrt{\left(\frac{\varepsilon_{M}^{2}}{2}+\lambda_{L}^{2}+\lambda_{R}^{2}\right)+\sqrt{\left(\frac{\varepsilon_{M}^{2}}{2}+\lambda_{L}^{2}+\lambda_{R}^{2}\right)^{2}-4\lambda_{L}^{2}\lambda_{R}^{2}}}\label{eq:epsilonQD}
\end{equation}
and
\begin{equation}
\epsilon_{\text{MZM}}^{\pm}=\pm\frac{1}{\sqrt{2}}\sqrt{\left(\frac{\varepsilon_{M}^{2}}{2}+\lambda_{L}^{2}+\lambda_{R}^{2}\right)-\sqrt{\left(\frac{\varepsilon_{M}^{2}}{2}+\lambda_{L}^{2}+\lambda_{R}^{2}\right)^{2}-4\lambda_{L}^{2}\lambda_{R}^{2}}}\label{eq:epsilonMZM}.
\end{equation}
\end{widetext}
For the case of MZMs with less degree of Majorana nonlocality, corresponding to Fig.~\ref{fig:SIAnalyticalSpectrum}(c) and Fig.~4(d) of the manuscript, $\lambda_{L},\lambda_{R}\gg \varepsilon_M$, and hence, from Eqs.~(\ref{eq:epsilonQD}) and (\ref{eq:epsilonMZM}) we obtain $\Omega_{M}=\sqrt{\epsilon_{\text{MZM}}^{\pm}/\epsilon_{\text{QD}}^{\pm}}=\sqrt{\lambda_{R}/\lambda_{L}}$ for the degree of Majorana nonlocality, which is exactly the same relation originally proposed by Prada et al~\cite{Prada2017}.


\providecommand{\noopsort}[1]{}\providecommand{\singleletter}[1]{#1}%

\medskip
\textbf{Author contributions}

LSR, VKK and IAS conceived the project. LSR carried out the calculations and plotted the figures, with contributions from VKK. LSR and IAS wrote the paper with contributions from ACS and VKK. All authors revised the manuscript.

\medskip
\textbf{Competing Interests}

The authors declare no competing interests.

\end{document}